 \documentclass[preprint,12p]{elsarticle}

\usepackage{amssymb}
\usepackage{lipsum}
\usepackage{soul}

\usepackage[utf8]{inputenc}

\usepackage{amsmath}

\usepackage{amsfonts}

\usepackage{amssymb}

\usepackage{graphicx}

\usepackage{booktabs}

\usepackage{xcolor}

\usepackage{array}

\usepackage{calc,ragged2e}

\usepackage{subcaption}

\usepackage{float}

\usepackage{tabu}

\usepackage[pdftex]{hyperref}

\usepackage{etoolbox}
\usepackage{multirow}
\usepackage{color}
\usepackage{makecell}

\definecolor{gris1}{gray}{0}

\definecolor{gris2}{gray}{0.45}

\definecolor{gris3}{gray}{0.6}

\definecolor{gris4}{gray}{0.75}

\definecolor{gris5}{gray}{0.9}

\makeatletter

\patchcmd{\@makecaption}

{\scshape}

{}

{}

{}

\journal{Egyptian Informatics Journal}

\begin{document}

\begin{frontmatter}


\title{Automatic Generation of Expert-Level Neuron Segmentation Masks from Fluorescence Microscopy Images for Non-Invasive Deep Learning Analysis of Phase-Contrast Images}

\author[uniovi_ee]{Gerard Villarroya-Piqué\corref{cor1}}
\ead{villarroyagerard@uniovi.es}

\author[uniovi_ee,uniovi_bme]{Víctor M. González}
\ead{vmsuarez@uniovi.es}

\author[uniovi_bio,uniovi_iuba]{Esther Serrano-Pertierra}
\ead{serranoesther@uniovi.es}

\author[uniovi_bio,uniovi_iuba]{M. Teresa Fernandez-Sanchez}
\ead{mfernandez@uniovi.es}

\author[uniovi_psy,uniovi_iuba]{Antonello Novelli}
\ead{anovelli@uniovi.es}

\author[uniovi_cs,uniovi_bme]{Angel Rio-Alvarez}
\ead{rioangel@uniovi.es}

\cortext[cor1]{Corresponding author}

\address[uniovi_ee]{Electrical Engineering Department. University of Oviedo, Asturias,  Spain}
\address[uniovi_cs]{Computer Sciences Department. University of Oviedo, Asturias,  Spain}
\address[uniovi_bme]{Biomedical Engineering Center (BME). University of Oviedo, Asturias, Spain}
\address[uniovi_bio]{Biochemistry and Molecular Biology Department, University of Oviedo, Asturias, Spain}
\address[uniovi_iuba]{University Institute of Biotechnology of Asturias (IUBA), University of Oviedo, Asturias, Spain}
\address[uniovi_psy]{Psychology Department, University of Oviedo, Asturias, Spain}

\begin{abstract}

\textbf{Background and Objective} 

Accurate segmentation of neurons in microscopy images of neuronal cultures is crucial for research on neurodegenerative diseases and neurotoxicity. Manual annotation of such images is time-consuming, subjective, and inconsistent across experts. Deep learning (DL) models offer an effective alternative, but require high-quality training datasets composed of microscopy images with accurately segmented neurons, typically created by experts.

Neuronal cultures can be imaged using either phase-contrast or fluorescence microscopy. While live neurons are easily detected in fluorescence images, this modality requires staining, which affects cell viability. Conversely, phase-contrast imaging is non-invasive but makes neuron identification more challenging.

\textbf{Methods}

In this work we present a robust computer vision pipeline for the automatic generation of expert-level neuron segmentation masks from fluorescence microscopy images. These masks are used to train DL models for neuron identification in phase-contrast images. Our method relies on paired fluorescence and phase-contrast images captured simultaneously at the same magnification and coordinates, ensuring alignment across modalities. Once trained, the DL models can segment neurons in phase-contrast images alone, eliminating the need for fluorescence staining.

\textbf{Results}

We validate our approach by comparing the automatically generated masks with expert annotations using multiple agreement metrics, including Intersection over Union, Accuracy, and Gwet’s AC1. Such metrics are calculated not only for the identification but also for the segmentation of instances.

\textbf{Conclusions}

This study presents a classical computer vision algorithm for generating neuron segmentation masks with reliable instance separation, which can be used for downstream applications such as non-invasive cell viability analyses by means of Convolutional Neural Networks. The results demonstrate that our method achieves expert-level performance, contributing to the field of neurobiological research.
\end{abstract}


\end{frontmatter}




\section{Introduction}\label{sec: intro}



Accurate identification and segmentation of neurons in microscope images of neuronal cultures are critical for research on neurodegenerative diseases and neurotoxicity. Traditionally, this task relies on manually counting live neurons, a process that is time-consuming, labor-intensive, and prone to variability across experts \cite{Gooliaff2018, Arimone2007}. The complexity of cell morphology, the presence of overlapping structures, and subtle viability cues further complicate this manual analysis.



Computer vision techniques, particularly those based on deep learning (DL), offer a promising alternative \cite{you2022multiscale, Chen2021, Hirapara2023} by enabling automated analysis of large image datasets. However, these models require extensive annotated training data, which is difficult to obtain in this domain due to the effort and subjectivity involved in manual labeling. Moreover, while unsupervised segmentation methods can efficiently cluster pixels into broad categories, they fall short in providing precise instance-level segmentation of neurons, a necessity for downstream tasks like cell tracking and viability assessment.



Microscopy images of neuronal cultures are typically acquired using two modalities: phase-contrast and fluorescence. Phase-contrast imaging is non-invasive \cite{Zernike1942} and thus suitable for longitudinal studies, but lacks the necessary contrast to reliably distinguish live neurons. In contrast, fluorescence microscopy enhances neuron visibility through staining—most commonly with fluorescein diacetate (FDA)—allowing accurate segmentation \cite{TesisCabrera}, albeit at the cost of cell viability. Consequently, fluorescence imaging is unsuitable for real-time monitoring or extended analysis of neuronal cultures.


Further complicating the task are the structural features of the culture itself: glial cells, neurites, and debris can obscure the presence of neurons. Viability assessments often depend on subtle morphological cues such as cell size, brightness, and structure integrity. Moreover, neurons frequently cluster, and individual segmentation becomes increasingly difficult, even for trained experts.


To overcome these limitations, various segmentation techniques have been proposed, including both supervised and unsupervised DL methods. While supervised models can offer high precision, they require large and diverse datasets tailored to specific imaging conditions. Unsupervised methods such as STEGO \cite{hamilton2022unsupervised} and CAUSE \cite{Seong_CAUSE}, have demonstrated outstanding performance on benchmarks like \cite{Cityscapes}. These methods excel at grouping pixels into broad categories without requiring labeled training data. Our task requires not only segmentation but also the precise separation of individual neurons. Methods like those discussed in \cite{Watson2023} and \cite{Wang2022} are typically used in applications where the individual instances of the members of a class are not relevant. Since our study aims to generate masks equivalent to expert annotations, these techniques are therefore unsuitable for our purpose. 

On the other hand, supervised DL methods require large and diverse datasets to avoid overfitting and to ensure robust feature representation \cite{Lones2021}. For fluorescence images, segmentation models create segmentation masks that indicate whether each pixel belongs to a neuron or the background. However, the key challenge is not just detecting neuron bodies but also ensuring their correct separation, especially when they cluster together. Even when leveraging transfer learning, models must be trained on highly specific datasets to learn this distinction effectively. Moreover, tools like Pyclesperanto-prototype \cite{Pyclesperanto_prototype} have introduced efficient GPU-based solutions for biomedical image processing, though they require careful tuning and do not address instance separation explicitly.

In this work, we propose a classical computer vision algorithm capable of automatically generating expert-level segmentation masks from fluorescence microscopy images obtained using the standard staining method \cite{Jones1985}. These masks can then be used to train DL models for non-invasive analysis of phase-contrast images. A key innovation of our approach is its ability to separate individual neurons—even in clustered regions—by combining brightness-based heuristics with shape and size filtering techniques. The algorithm leverages paired fluorescence and phase-contrast images taken at identical magnification and coordinates, ensuring one-to-one correspondence across modalities.

The algorithm processes fluorescence images using a series of low-level transformations, filters, and smoothing techniques to distinguish between foreground (live neurons -\textit{on}-) and background (neurites, glia, debris, and dead neurons -\textit{off}-), in order to automatically generate segmentation masks. Special attention is paid to instance-level segmentation, enabling the model to separate closely packed neurons and produce masks also suitable for object detection tasks. The foreground-to-background pixel ratio (1:10) also influences performance metrics \cite{FERRI200927, SunImbalanced2009}, requiring careful evaluation.


To validate the method, we compare the automatically generated masks with those produced by expert annotators for the same dataset of images. Particular emphasis is placed on metrics that are robust to label imbalance and inter-rater variability, as there is no universal agreement on the best index to measure it \cite{Gisev2013, Feinstein1990, Zhao2022}. Indices like MCC \cite{Chicco2021} and Gwet's AC1 \cite{Gwet2008, MansillaEpilepsyExpertAgreement} address the paradoxes in Cohen's Kappa, but are not universally applicable. The choice depends on the characteristics of the study \cite{Vach2023}, described in Section \ref{sec: means}.


Evaluation is conducted using both pixel-level metrics (e.g., Gwet’s AC1 and Accuracy, claimed as more stable \cite{Zhao2022}), where confusion matrices are used to compare the classification of individual pixels, and instance-level metrics (e.g., Intersection over Union, IoU), to measure out segmentation quality. All these indices are not supposed to be substitute for other ones \cite{Vach2023}.

\section{Methods}\label{sec: means}

\subsection{Primary neuron cultures}\label{subsec:PrimaryNeuronCultures}
The appearance of neurons under a microscope varies depending on their type and the imaging technique used. To develop a robust segmentation model, it is essential to acquire images with consistent characteristics. This section details the imaging techniques and dataset used in this study, ensuring reproducibility and compatibility for future dataset expansion.

Primary neuron cultures are obtained from the cerebellum of 6–7-day-old mice, where neurons are not fully developed yet. Cultures of cerebellum neurons show consistent homogeneity in cell shape and size \cite{Shabanipour2019}. The dataset consists of 31 images. These images show specific culture areas taken using two techniques: phase-contrast and fluorescence staining. Phase-contrast imaging accentuates cellular structures, making cells, glial cells, neurites, and debris clearly visible. They generate complex images where cells obfuscate each other, rendering identification by inspection unfeasible. For this reason, images are taken using fluorescence, chemically staining the culture. The staining technique leverages the ability of live cells to absorb and metabolize specific chemicals. Fluorescein diacetate (FDA) is a membrane-permeable compound that, once inside the cell, is hydrolyzed by intracellular esterases to produce fluorescein. This reaction only occurs in live neurons with intact membranes, resulting them to fluoresce under the microscope \cite{Jones1985}. As a result, fluorescence images provide a reliable ground truth for neuron identification, as only viable cells are stained \cite{Nozhat2022}.

The main drawback of this method is that, to get the cells to incorporate the fluorescein diacetate, the membrane must be made permeable. It damages the cell membrane enough that the culture will not survive for long. Thus, monitoring a culture over time using this technique is not possible. A standard phase-contrast microscope and an Olympus IMT-2 digital camera with a Sony SPT-M308CE video camera were used to take the images.

Not all images were suitable for analysis due to limitations inherent to the staining technique. Over time, fluorescein diffuses out the cells, also staining the background and thus reducing the contrast between FG and BG. This effect compromises the effectiveness of segmentation, making it difficult to distinguish neurons from the surrounding medium. After consulting with experts, four images were identified as lacking the minimum contrast required for meaningful analysis and were excluded from the dataset, leaving a final set of 27 images. A preliminary approximation of the number of instances in the dataset results in the expected number of neurons to be close to 3,000 instances. It has been determined by manually counting the neurons present in 2 images that the experts considered to be representative, totaling approximately 110 instances per image.

\subsection{Annotation software}\label{subsec:Krita}
The fluorescence images were meticulously annotated by three experts to establish a reliable ground truth for neuron counting. The diversity of the images in terms of neuronal density and morphology allows for a comprehensive evaluation of the algorithm's performance under different conditions. A protocol was established to minimize tool-related inconsistencies. The open-source raster graphics editor Krita (Disponible en: \url{https://krita.org/en/}) was used by the experts to to manually draw the binary masks. Once installed, several settings were adjusted to draw the binary masks properly.

The experts agreed on settings ensuring that pixels were strictly black or white, without anti-aliasing on the edges.

The following procedure was adopted to create binary masks from fluorescence images:

\begin{enumerate}
    \item Import the fluorescence image to Krita.
    \item Add a new paint layer via the \textit{Layers} menu. This layer will serve as the binary mask.
    \begin{enumerate}
        \item Use the \textit{Fill Tool} to paint the entire mask layer black, establishing the background.
        \item Reduce the opacity of the mask layer to visualize the underlying fluorescence image.
    \end{enumerate}
    \item With the mask layer selected, use the \textit{Freehand Selection Tool} to delineate the outline of each individual neuron or cluster of neurons. Fill the selected region with white color using the \textit{Fill Tool}.
    \begin{enumerate}
        \item In the case of clusters of neurons, manually draw black lines to separate adjacent somas.
    \end{enumerate}
    \item Repeat the previous step for all neurons in the image.
    \item Restore the mask layer’s opacity to 100\%.
    \item Flatten the layers and export the final binary mask as a \textit{.tiff} file.
\end{enumerate}

This protocol for the annotation of the images ensures the creation of high-quality binary masks that serve as ground truth reference for the evaluation of the performance of the segmentation algorithm.

\subsection{Generation of binary segmentation masks from fluorescence images }\label{subsec: genmasks}

This algorithm has two main goals. On the one hand, to generate binary segmentation masks from fluorescence images. And on the other hand, to accurately identify and separate individual neurons even when they overlap or form clusters. Unlike simple thresholding techniques that merely binarize an image, this method extracts neuron instances while filtering out background noise, glial cells, and neurites. The process starts by ensuring that the input image is in the correct format. It should contain a single channel encoding pixel brightness as an integer between 0 and 255. Fluorescence images exhibit varying brightness levels, making an approach based on a fixed value to threshold the image not accurate. Instead of using a single mean pixel value, the algorithm dynamically adjusts the thresholding value to enhance neuron visibility across different images.

The foreground is understood as all the pixels that are part of any neuron. To locate neurons, all local brightness maxima are identified. The images exhibit multiple local maxima that do not belong to the foreground, which is where neurons are. To address this, a standard procedure to make a rough separation between the foreground and the background, consisting in applying a Gaussian blur followed by Otsu's thresholding \cite{Otsu1979} is performed. The candidate locations are then selected based on whether they lie within the foreground. This step is quite important as the algorithm will search for neurons only in the locations outputted.

At this stage, the algorithm has pinpointed potential neuron locations. The next stage of the algorithm separates actual neurons from other cells and background noise. To do this, the algorithm relies on the consistent size and shape of live neurons within this cellular type. For each candidate location, it isolates the area by drawing a fixed-size box around it. Within these boxes, contours are detected and fitted into ellipses. Size filters exclude irrelevant shapes, and the validated ellipses are filled and drawn as white onto a black image. This image becomes the mask, where each ellipse marks a neuron’s location and acts as a close approximation of its shape. By performing a binary AND operation with the base foreground, the shapes of the ellipses fit the neurons perfectly.

Now, the algorithm tackles the separation of individual neurons from one another, with emphasis on the clustered neurons. It combines the cropped mask with the earlier candidate locations using a powerful function from the \textit{Pyclesperanto} library. This function assigns unique colors to each instance, allowing clear differentiation. Black lines are drawn wherever the colors change, resulting in a mask that separates each neuron into distinct regions. While the initial mask successfully delineates neuron instances, it may contain rough edges or small artifacts. To refine the segmentation, the algorithm applies morphological operations such as erosion and dilation. This process smooths neuron contours, eliminates small artifacts, and ensures that the masks are clean and reliable for eventual use in training deep learning models.

The final output is a binary mask that accurately encodes the shape of neuron somas, effectively separating them from glial cells, neurites and other neurons, which appear bright in the original fluorescence images. By leveraging instance separation techniques, the algorithm successfully distinguishes clustered neurons, ensuring precise segmentation suitable for downstream analysis.

\subsubsection{Preprocessing}
For the algorithm to perform optimally, images should have consistent brightness and contrast. Areas within the same image with different intrinsic background intensity skew the dynamic brightness computations. This step makes the posterior analysis more reliable. It consists of two interlocked steps that perform heavy lifting for the algorithm. Both steps aim to smooth out the background variability and to increment the contrast at the edges of the neurons:
\begin{enumerate}
    \item Gradient computation and binary thresholding.
    \item Computation of the Gaussian blur and subtraction.
\end{enumerate}

These steps work in tandem; the output of step 1 serves as a mask to select the areas where to apply step 2. The first part of the pipeline, that is step 1, computes a low-intensity Gaussian blur of the image to remove the fine details that make differential calculus meaningless. Differentiation of noisy data is known to amplify the noise, and techniques have arisen to tackle the problem \cite{DiffOfNoisy2014}. Then, taking the gradient, it returns an image that has high values in the regions close to the edges of the neurons. That is, it marks the areas of the image where there are intense brightness changes, $D(I)$. This image holds accurate information about the locations of the edges of neurons and clusters and can be used as a mask to select the areas where step 2 should be applied. 

A parallel operation, complementary to the previous one, enhances contrast near neuron edges by subtracting the blurred image to itself. After applying a low-intensity Gaussian blur, areas with homogeneous pixel values will not change. The pixels close to the edges change. After subtracting the two images, the result has two distinct areas: 1) constant brightness areas will have values close to zero; 2) close to the edges, the side with lower brightness will result in a negative value and the side with high brightness will result in a positive value. After setting the type of values to be represented by an unsigned integer, the negative values wrap around, resulting in a very high value. This constraint produces an interesting image where pixels immediately outside neurons and clusters take a value close to the maximum. The values of the other pixels are much lower in comparison. The image obtained, $B(I)$, is represented by equation \ref{eq:1}, where $I$ is the fluorescence image, $Blurred(I)$ indicates the image after applying the Gaussian blur, and \textit{uint8} stands for ``unsigned integer of 8 bits", thus storing a positive integer from 0 to 255, both included. 

\begin{equation} \label{eq:1}
    B(I) = I - Blurred(I)\textit{ ,   as uint8}
\end{equation}

Once this preprocessing is over, the background can exhibit high-valued pixels. We then compute the element-wise minimum between this inverted image and the original fluorescence image, shown in equation \ref{eq:2}. This operation preserves neuron body intensities while enhancing contrast at their edges. The result is a refined image where neuron's boundaries are more pronounced and background interference is minimized. The result of step 1, $D(I)$, functions as a way of masking out the background, so all pixels outside the relevant areas are set directly to zero.

\begin{equation} \label{eq:2}
    \textit{Treated Image} = \min(255 - B(I),I)\textit{, where $D(I) \neq 0$ }
\end{equation}

At the end of this process, the resulting image retains the key structural features of the original while exhibiting enhanced edge contrast and reduced background noise. This preprocessed image serves as input for the next stage of the algorithm, where the foreground and background are precisely differentiated.

\subsubsection{Foreground and background separation}

The algorithm processes single-channel grayscale images with pixel values in the range [0, 255]. After preprocessing, the images are standardized in format, but variations in brightness remain, both across images and within individual samples. This variability hinders the separation of foreground (FG) and background (BG), especially in regions where neurons appear faint.

Pixels not associated with any stained structure are considered background (BG), while all other pixels constitute the foreground (FG). However, the stain highlights all components of living cells —not just neuron's soma— including neurites and glial cells. Therefore, a reliable FG/BG separation is essential before isolating neuron's bodies. A simple threshold using the mean pixel intensity $\mu$ is insufficient because BG pixels dominate the image, skewing $\mu$ towards the background. To correct this, a positive offset called \textit{Tolerance} ($T$) is added, producing an adjusted threshold $\mu' = \mu + T$. This correction allows for more accurate separation by excluding weakly stained or noisy regions.

\begin{figure}[hbt]
	\centering
	\includegraphics[width=\columnwidth]{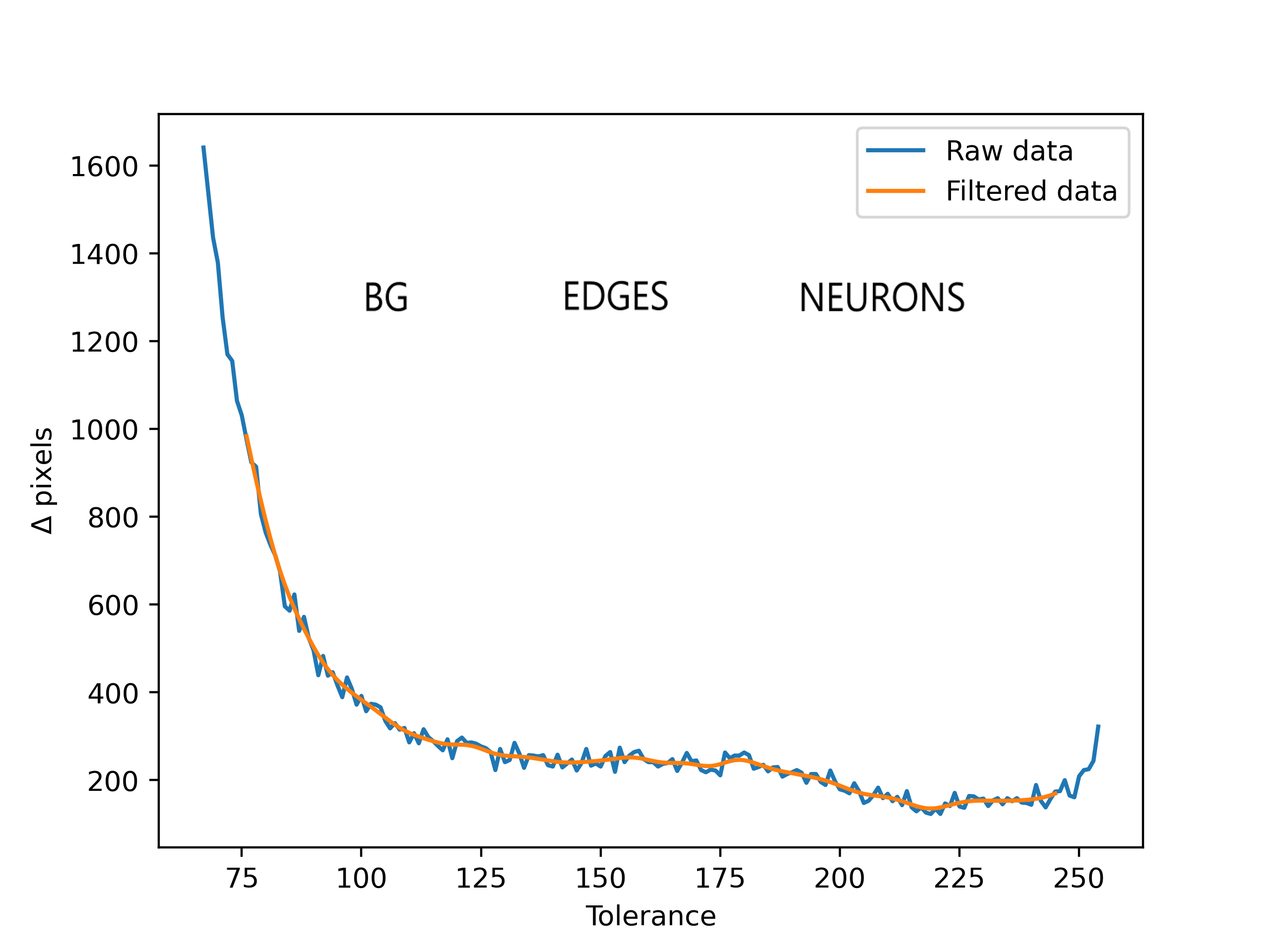}
	\caption{The filtered data is calculated by a convolution with a Gaussian kernel. As the \textit{Tolerance} increases, 3 regions can be identified. The limit values between the three regions are not clearly defined.}
	\label{fig: T_Regions}
\end{figure}

The algorithm scans through a range of $T$ values, identifies the FG and keeps track of the number of pixels that change from FG to BG as $T$ increases. The value at $T$ is the difference between the number of pixels in the FG calculated at $T$ and $T-1$. Figure \ref{fig: T_Regions} shows three distinct regions.

For low values of $T$, the number of pixels that change from FG to BG is high and quickly diminishes. It is due to $\mu$ being low; the FG includes pixels that are not stained. Ideally, those pixels should belong to the BG. The region at low $T$ is the \textit{BG region}. As $T$ increases, the number of pixels that change from FG to BG becomes constant. This region corresponds to threshold values that coincide with the values at the edges of the neurons. It is the \textit{EDGES region}. The width of this area correlates to the contrast between a neuron and the BG. Higher contrast would produce a broader region. The \textit{NEURONS region} is made up of high $T$ values. In this right-most region $\Delta pixels$ increases as the soma of the neurons and glial cells change into part of the BG.

This allows us to apply the algorithm at specific intervals for $T$, reducing the values searched and making the algorithm faster. Only the \textit{EDGES region} is relevant in the search. The lower limit of the interval is computed by taking the first derivative of the filtered data. The first value where the derivative gets close to zero is called $T_{0}$. $T$ starts the search at $T_{start} = T_{0} - 10$, ensuring the algorithm searches all values in the \textit{EDGES region}.

The algorithm is executed for consecutive values of $T$, and the number of neurons detected in the final mask for the current $T$ is stored. Once this value is lower than $80\%$ of the initial value, the search stops and the algorithm returns the data corresponding to the $T$ that produced the last maxima in the neuron count.

This step ensures that subsequent stages operate on a high-quality foreground, where neuron bodies are well-defined and background interference is minimized. As the effectiveness of the entire segmentation pipeline depends on this initial separation, careful optimization of $T$ is essential.

\subsubsection{Finding the candidate locations}
Live neurons typically appear as bright spots in the FG due to the fluorescence stain. Other elements of the image that are also alive show similar values for the brightness intensity, but differ in intrinsic characteristics such as shape and size. The algorithm identifies brightness maxima in the image to create a map of potential neuron locations. Structures like the connections between neurons and the projections emitted by the glial cells don't appear as local maxima in brightness. Clusters of neurons pose a significant and ubiquitous challenge, as overlapping neurons may obscure individual maxima. While some neurons in a cluster may still exhibit local maxima, their perimeters are not always well-defined by the BG. This step is particularly crucial because neurons missed at this stage will not be detected in later phases. To minimize false negatives, candidate locations are identified in excess, allowing subsequent filtering steps to remove non-neuronal structures.

To ensure accurate maxima detection, a Gaussian blur is applied beforehand. This step reduces noise and smooths fine details that might otherwise introduce not genuine maxima. The \textit{Pyclesperanto} library, available at \cite{Pyclesperanto_prototype}, provides optimized functions for this operation. This library also provides functions to detect maxima in an image. The function that detects maxima is applied to every pixel in the blurred image. The return value is $1$ if there are no higher-valued pixels within a given distance, and $0$ otherwise. Mathematically, it takes the form given by equation \ref{eq: Maxima}.

\begin{equation}\label{eq: Maxima}
\text{DetectMaxima}(\text{\textit{pixel}}) = \begin{cases} 
      1 & \text{\textit{pixel}} > \text{\textit{surrounding pixels}} \\
      0 & \text{\textit{otherwise}}
   \end{cases}
\end{equation}

After applying the function to every pixel, its output is a binary image that marks the locations where maxima have been identified ($L$). The previously-applied blur allows the function to only consider the closest pixels for the calculation. The blurred image has information about the general structure of the features of the image, where the variation given by the noise has been reduced. Having multiple maxima identified for a single neuron is correct because the later steps of the algorithm will remove the redundancy.

In a parallel computation to the detection of maxima, another lower-intensity Gaussian Blur is applied to the image to remove unwanted noise. The specific combination of values used is chosen after testing the range of possible options. The new image is used to obtain a separation between the FG and the BG that will be used to realize which of the detected maxima are truly part of the FG. The well established method of Otsu's thresholding \cite{Otsu1979} performs the separation the algorithm needs, obtaining a mask for the FG ($FG'$). The \textit{Pyclesperanto} library provides an optimized implementation of Otsu's method, leveraging \textit{Scikit-image} for efficient processing.

The algorithm has now generated two new images that hold information. One image ($L$) indicates where local maxima have been identified in the FG. Another image ($FG'$) holds information about the areas where maxima are relevant. The process guarantees that the image encodes the areas of high brightness, which encode the neurons. After performing a binary AND on $FG'$ and $L$, the algorithm finishes the identification of the candidate locations. By doing a logic AND, only the pixels holding the value 1 in $L$ are affected. When the maxima in $L$ is located in the background, $FG'$ has a value of 0, so the maxima in $L$ is discarded. Only maxima in the foreground go thorough this filter.

\subsubsection{Size and shape analysis; a method with ellipses}

Shape analysis plays a crucial role in distinguishing neurons from other cellular structures and ensuring accurate instance segmentation. While neurons in primary cultures exhibit relatively homogeneous morphology, they frequently cluster together, making individual segmentation challenging. Their elliptical shape provides a reliable feature for differentiation, as neurons tend to have low eccentricity compared to glial cells and neurites. This step refines neuron detection by filtering out candidates that do not meet expected size and shape criteria.

In order to identify the shape of each relevant location, the algorithm loops over the coordinates encoded in $L$ to analyze whether a neuron is in the location or not. For each one, it isolates the location from the surrounding area by cropping a box of fixed size. The size is chosen so it ensures the whole soma of the neuron is contained within, but small enough to minimize interferences from unrelated areas of the image. In each isolated region, it identifies the underlying shapes using \textit{find\_contours} from the \textit{OpenCV} library, a robust, optimized library for computer vision integratied in Python. Each contour detected results in an ellipse being fitted. Several ellipses are generated for each location and are filtered out based on size and eccentricity. A challenge arises when a cluster is treated.

Glial cells reach this point of the algorithm as neuron candidates since they exhibit similar features. Making a distinction between glial cells and clusters of neurons can be challenging. Glial cells' size is greater than that of neurons. This type of cells appear as wide areas of homogeneous intensity, such that they don't show a closely packed distribution of maxima in brightness but a single broad maxima. However, clusters of neurons exhibit several maxima close together. The algorithm applies a shape and size filter to the generated ellipses.

Some neurites can enter this stage of the analysis, although they don't consistently show maxima in brightness. Their shape is a differentiating factor since it has a much higher eccentricity than the average neuron.

The set of filtering conditions applied to refine the candidate ellipses is comprised of:
\begin{enumerate}
    \item Size constraints: Ellipses with axes exceeding a predefined range are discarded, as excessively large ellipses likely correspond to glial cells or neuron clusters.
    \item Eccentricity threshold: Ellipses with a major-to-minor axis ratio beyond a given limit are removed, as these typically correspond to neurites.
    \item Aspect ratio correction: Only ellipses approximating a circular shape with biologically plausible dimensions are retained.
\end{enumerate}

After applying these filters, the remaining ellipses are drawn onto a black image, forming a preliminary segmentation mask referred to as the \textit{proto-mask}. The proto-mask serves as an initial approximation of neuron locations, encoding the detected ellipses as white regions on a black background. Since neurons may be detected multiple times due to overlapping candidates, redundant markings have no effect, as white pixels are simply drawn on top of existing white regions. The neuron instances in the image aren't segmented in the proto-mask yet.

\begin{figure}[ht]  
\begin{subfigure}{0.48\textwidth}
\includegraphics[width=\linewidth]{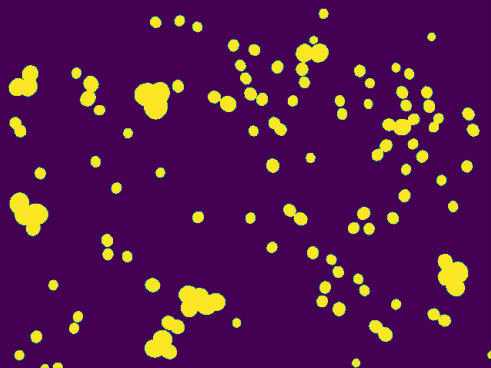}
\caption{}
\end{subfigure}
\hfill 
\begin{subfigure}{0.48\textwidth}
\includegraphics[width=\linewidth]{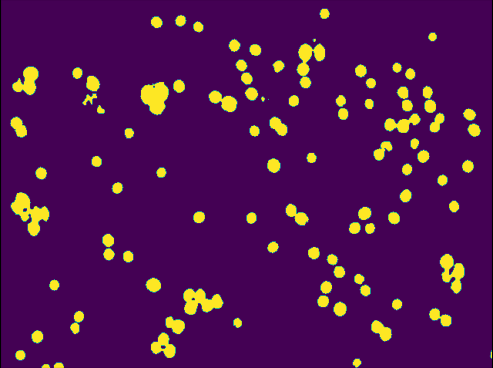}
\caption{}
\end{subfigure}
\caption{Process to obtain the mask previous to instance separation. (a) shows the mask after the ellipses have been drawn to it. Since neurons are not perfectly elliptic, ellipses are fitted to their shapes, as shown in (b).}
\label{fig: BinaryAnd}
\end{figure}

Since neurons are not perfect ellipses, the proto-mask undergoes a refinement step to better fit their actual morphology. A binary AND operation is performed between the proto-mask and the thresholded original image $FG'$, ensuring that the final segmentation adheres more closely to the neuron's contours. This process is illustrated in Figure \ref{fig: BinaryAnd}, where (a) shows the initial proto-mask and (b) depicts the refined mask after morphological adjustment.

\subsubsection{Neuron separation and edge delimitation}

\begin{figure}[ht]  
\begin{subfigure}{0.48\textwidth}
\includegraphics[width=\linewidth]{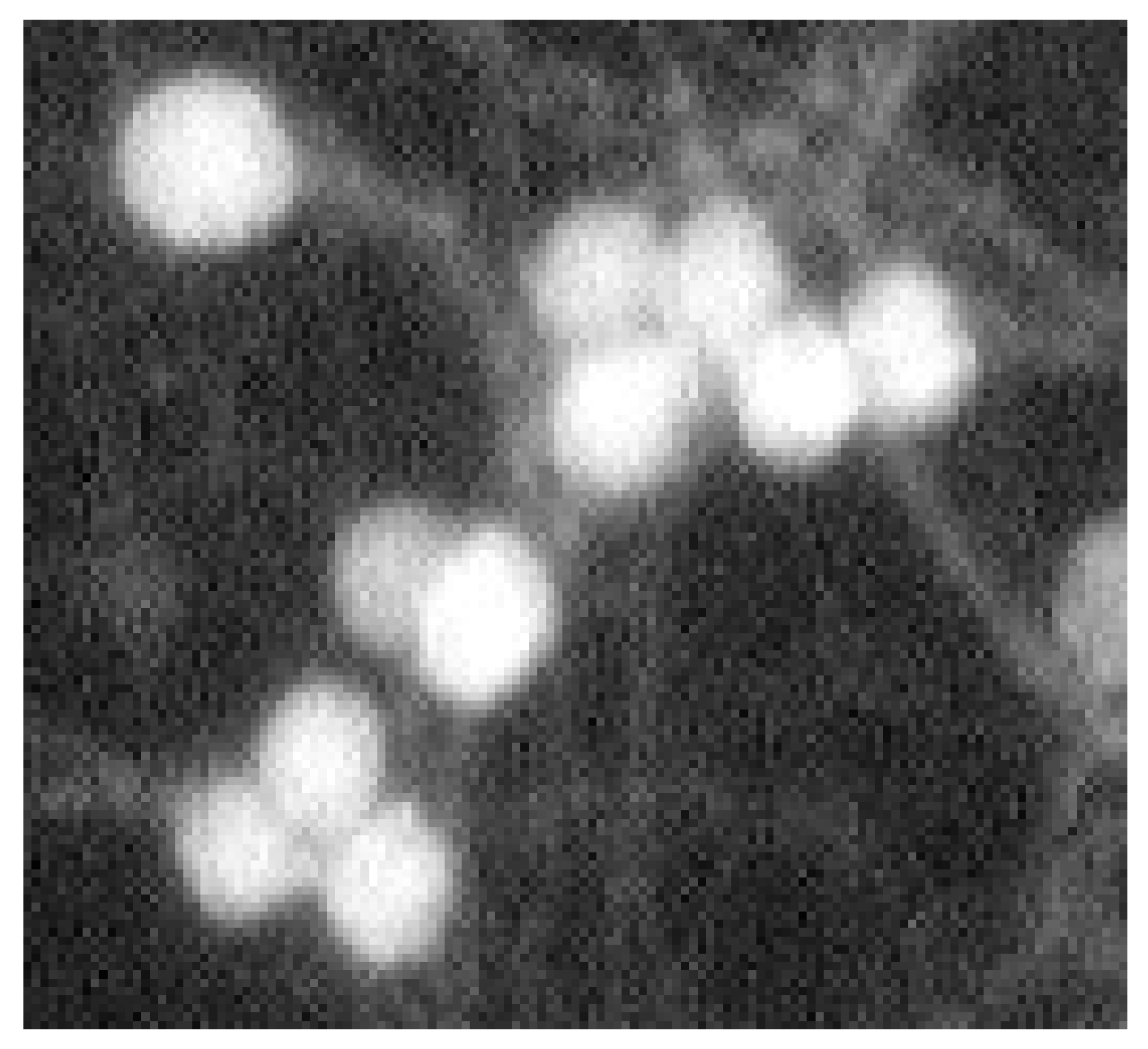}
\caption{Original image}
\end{subfigure}
\hfill 
\begin{subfigure}{0.48\textwidth}
\includegraphics[width=\linewidth]{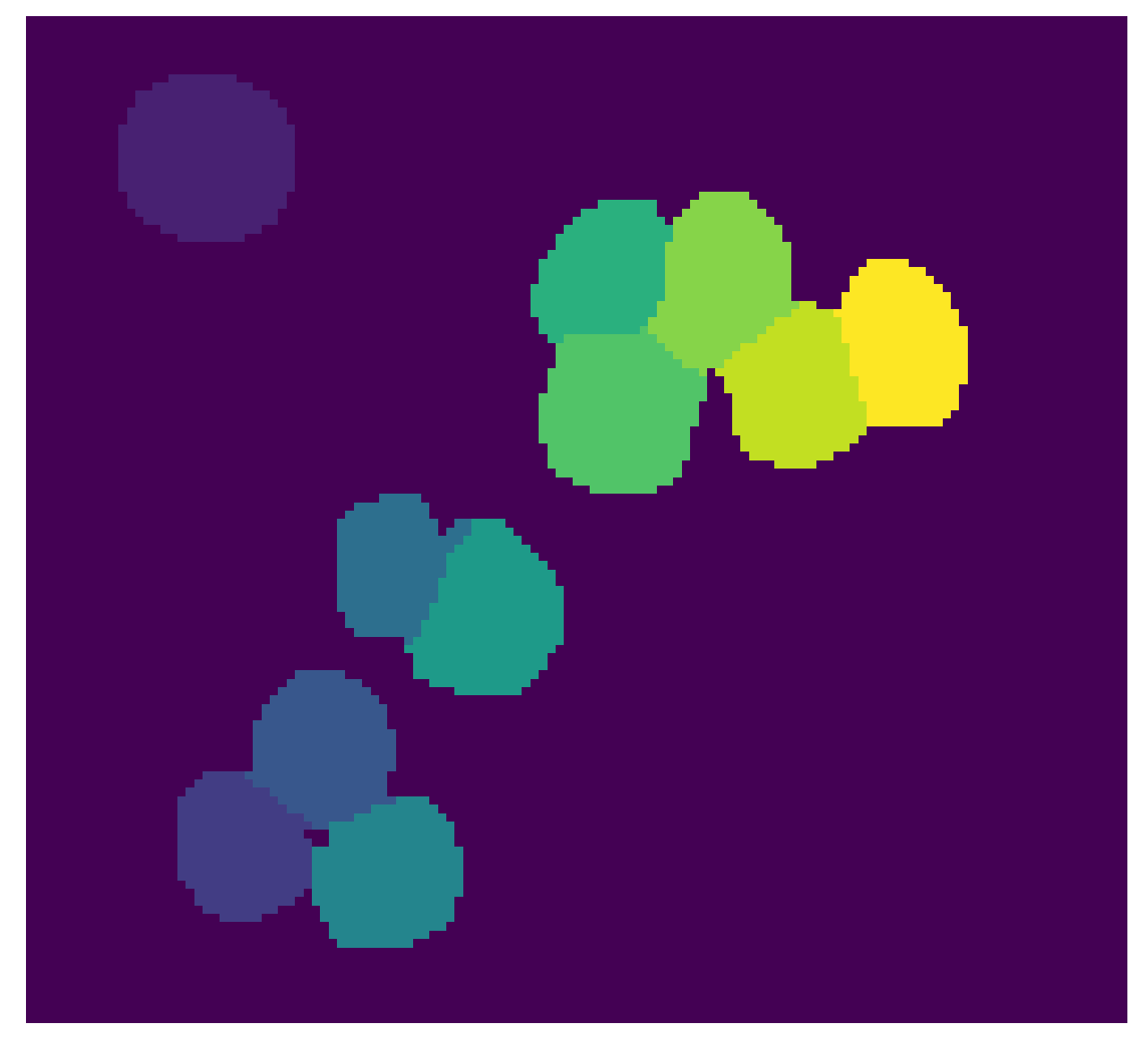}
\caption{Image $V$}
\end{subfigure}

\bigskip  
\begin{subfigure}{0.48\textwidth}
\includegraphics[width=\linewidth]{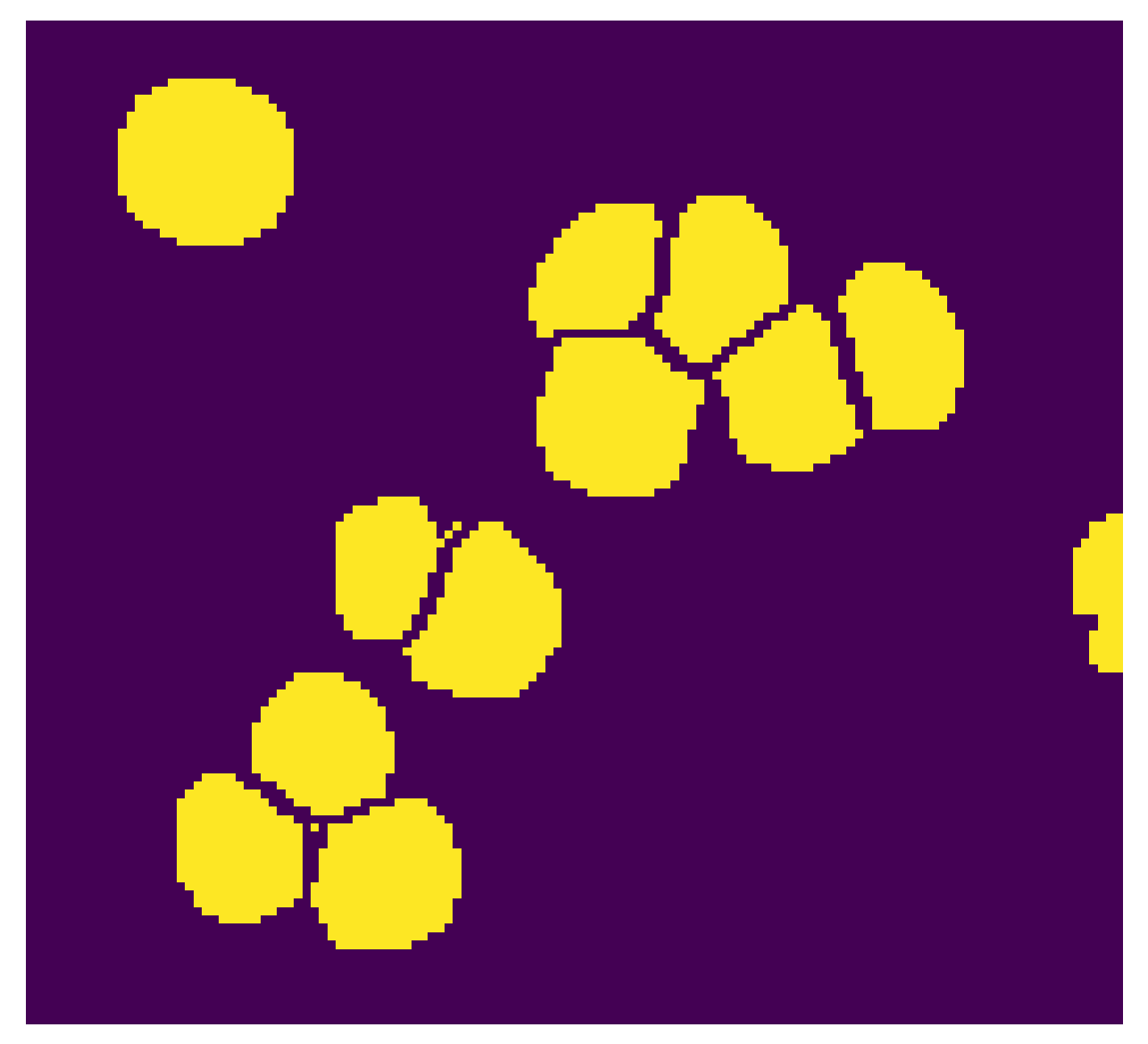}
\caption{Pixels in $\partial V$ have been erased}
\end{subfigure}
\hfill 
\begin{subfigure}{0.48\textwidth}
\includegraphics[width=\linewidth]{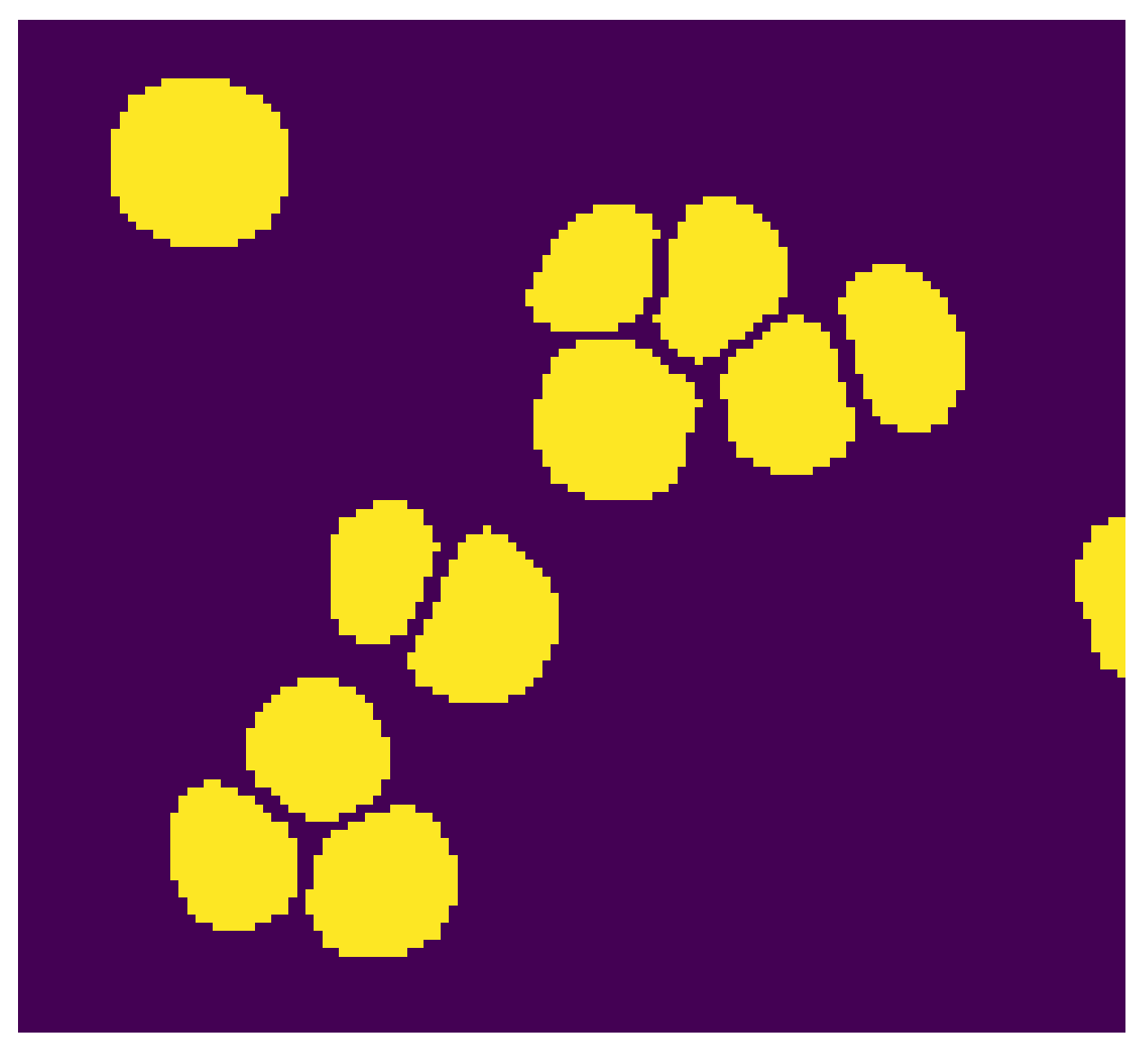}
\caption{Cleaned final mask after smoothing}
\end{subfigure}

\caption{Steps involved in the separation of neurons.} 
\label{fig: NeuronSeparation}
\end{figure}

Identifying the locations of neurons in the proto-mask isn't adequate because it's a binary image; it lacks local features. The image $L$ is better suited for this job, and has already been computed. It encodes the locations where a neuron will be. The algorithm employs the Voronoi labeling technique, which assigns unique labels to each neuron candidate based on local maxima detected in the previous step. The implementation of such process is obtained from \textit{Pyclesperanto}, where octagons are dilated at each location in $L$ until they touch another growing shape or leave the proto-mask. The function's output is an image where each polygon instance is labeled differently, denoted as $V$. Important pixels in $V$ mark the boundary between two different labeled areas. In the labeled image $V$, the boundaries between distinct neuron regions hold special significance. These boundaries, denoted as $\partial V$, define the separation between adjacent neurons. The algorithm will ensure that these borders are excluded from the final mask, preventing merged instances. The pixels in $\partial V$ are identified as those that transition between different nonzero labels, as illustrated in Figure \ref{fig: NeuronSeparation}. The boundary is only relevant between non-background labels. The image in the bottom right corner shows the final result after the next subsection \ref{subsec: smooth}.

To enforce proper neuron separation, the algorithm iterates through the proto-mask and removes any pixel that belongs to $\partial V$. This transformation can be formalized in equation \ref{eq: voronoi_sep} and Figure \ref{fig: NeuronSeparation}, where pixels labeled as 0 correspond to the background. The algorithm passes through all the pixels that are non-zero in the proto-mask and sets those that lie on the boundary to zero. This ensures that neurons remain distinct, preventing fused instances in the final segmentation mask.

\bigskip
\begin{equation}\label{eq: voronoi_sep}
  \begin{gathered}
    \text{InstanceSeparation}(pixel = 0) = 0 \\
    \text{InstanceSeparation}(pixel = 1) = \begin{cases} 
      0 & \text{\textit{pixel $\in \partial V$}} \\
      1 & otherwise
   \end{cases}
  \end{gathered}
\end{equation}
\bigskip 

The original images will have neurons that appear at the margins. All neuron bodies that aren't within the image mustn't be accounted for in the mask. Neurons with straight edges don't represent the full instance of a neuron. These partial neurons exhibit unnatural straight edges, which do not correspond to real biological structures and thus have to be excluded. The algorithm detects abnormal linear edges in the image and removes them. The analysis of the image's border is not enough because it can appear padded. 

The algorithm analyzes each border of the image independently. Taking the left edge as an example, it identifies the first column containing a neuron pixel and counts the number of consecutive pixels forming a contiguous block. If this block exceeds a predefined threshold, the entire neuron instance is removed from the mask. The threshold is determined empirically by comparing the distribution of neuron sizes in the dataset. Based on this analysis, a value of 9 pixels was chosen as the maximum admitted block size before removal.

\subsubsection{Smoothing} \label{subsec: smooth}

Before generating the final mask, the algorithm applies a smoothing step to correct artifacts and refine neuron's boundaries. The segmentation masks obtained through the AND operation often display jagged edges caused by pixel-wise thresholding, which can introduce inconsistencies. Additionally, minor imperfections in the overlap between the images can produce small pixel islands that do not correspond to actual neurons. These artifacts must be removed to ensure that the masks accurately represent neuron morphology and, thus, will be suitable for eventually training deep learning models. Repeatedly eroding and dilating the mask removes the jagged imperfections. The order is critical since dilating first would fill the separation between each neuron, and the image would lose information. By performing erosion first, small artifacts and isolated pixels are removed, ensuring that only well-defined neuron's structures remain. The subsequent dilation then restores neuron's shapes while maintaining clear separations between instances.

Erode and dilate are two complementary processes that modify the state of a pixel based on how many pixels around them are part of a neuron (\textit{on}) or part of the background (\textit{off}). Erode sets to \textit{off} those pixels with more than a predefined limit number of \textit{off} pixels around them. Dilate sets \textit{on} pixels surrounded by more \textit{on} pixels than a certain amount. 

\subsection{Metrics}\label{subsec: Metrics}

To evaluate the performance of the algorithm, the generated masks are compared against the expert's annotations. Since the masks serve both segmentation and object detection purposes, the comparison is carried out at two levels: pixel-level  (for segmentation quality) and instance-level (for object identification). Each level requires a different set of evaluation metrics.

At pixel level, the evaluation treats each image as a collection of binary classification problems, where each pixel must be classified as either belonging to a neuron or not. This allows the construction of a confusion matrix for each pairwise comparison. The algorithm is considered one of the raters and is compared against each expert.


\begin{table}[ht] \label{tab: CM_scheme}
\begin{center}
\tabulinesep=2mm
\begin{tabu} {l|l|c|c|c}
\multicolumn{2}{c}{}&\multicolumn{2}{c}{Expert}&\\
\cline{3-4}
\multicolumn{2}{c|}{}&Neuron&Not neuron\\
\cline{2-4}
\multirow{2}{*}{Algorithm}& Neuron & $TP$ & $FP$\\
\cline{2-4}
& Not Neuron & $FN$ & $TN$\\
\cline{2-4}
\end{tabu}
\caption {Confusion matrix for pixel-level evaluation.}
\end{center}
\end{table}

Based on the literature discussed in Section \ref{sec: intro}, the selected pixel-level evaluation metrics include the Matthews Correlation Coefficient (MCC), Jaccard Index, and a precision–recall pair. The MCC is a balanced metric suitable for imbalanced datasets, as it takes all cells of the confusion matrix into account. It is considered more robust than Cohen’s Kappa in certain cases \cite{Chicco2021}. Its formula is shown in equation \ref{eq: MCC}.

\begin{equation} \label{eq: MCC}
    MCC = \frac{TP\cdot TN - FP\cdot FN}{\sqrt{(TP+FP)(TP+FN)(TN+FP)(TN+FN)}}
\end{equation}

Accuracy is not used in this analysis due to the strong class imbalance: most pixels belong to the background class, which can inflate the metric despite poor segmentation \cite{Lazlo_Datasets}. Gwett's AC1 has been calculated instead to yield results analogous to those of the Accuracy, with similar stability features. The Jaccard index is a standard measure to compare similarities in segmentation tasks \cite{Eelbode2020} \cite{Ronneberger2015}. In the literature, it also appears under the name of \textit{Intersection over Union (IoU)}. This index is calculated both in a per-instance and global way. The global calculation yields a value for the image as a whole, applying the formula to the confusion matrix obtained for the whole image.

Corresponding neurons in each mask must be isolated prior to analysis. The analysis sets all pixels in the vicinity of the neurons that belong to other instances to zero. The IoU is then computed for each neuron instance using the formula. It allows us to confirm if a specific neuron was identified accurately. We have metrics at a pixel level for segmentation purposes and at instance level for object detection.

 At the instance level, the threshold for a positive detection is set at 0.3, following previous works \cite{Abdurahman2021, Hong2019}. While 0.5 is the standard threshold in object detection, a lower threshold is appropriate here, as the primary objective is location-level detection of neurons. Pixel-level precision is evaluated separately. IoU is defined in equation \ref{eq: iou}.

\begin{equation} \label{eq: iou}
    IoU = \frac{TP}{TP + FP + FN}.
\end{equation}

The metrics used have an interpretation, and can be translated into a qualitative scale. In the implementation of the \textit{pycm} library, an interpretation scale for most of these indices is provided \cite{Haghighi2018}. The MCC metric is considered to imply strong agreement if it is higher than $0.7$, and very strong if it exceeds $0.9$.

Precision and Recall are two fundamental metrics that give meaningful information when used in pairs. Precision measures the ratio of positive identifications within the total identifications, while Recall indicates how many of the actually positive elements have been retrieved. 
\begin{equation}\begin{gathered}
     P = \frac{TP}{TP + FP}\\R = \frac{TP}{TP + FN}
\end{gathered}\end{equation}

They can be analyzed as a pair in a Precision-Recall graph or combined in the F-measure, implemented as the F1 index. At a pixel level, F1 is not recommended due to class imbalance \cite{Brabec2020}. Nevertheless, instance identification doesn't share the same class-imbalance characteristics to the pixel domain. In this context, the F1 metric is a suitable way of implementing the information provided by $P$ and $R$ \cite{Taha2015}.

The F1 metric will give us a metric that balances the importance of $P$ and $R$, through their harmonic mean. The practical interpretation of this metric is disputed and it is possible to transform the F1 metric into another metric with direct interpretation, given by \cite{Hand2021}. The result is the Jaccard coefficient. This represents the fraction of correctly identified neurons relative to the total number of identified instances..

\begin{equation} \begin{gathered} \label{eq: F1-Jac}
F1 = 2\frac{P\cdot R}{P+R} \\
Jac = \frac{F}{2-F} = \frac{TP}{TP + FP + FN}
\end{gathered} \end{equation}

Note that equation \ref{eq: F1-Jac} is not the same as equation \ref{eq: iou}. IoU is a metric calculated from the confusion matrix at a pixel level, and its result is the global similarity at a pixel level. The $Jac$ coefficient is computed from the confusion matrix obtained at instance level.

\section{Results}\label{sec: results}

As discussed in the previous section concerning the metrics, analysis has to be performed at several levels. The key characteristic of this method is that it is designed to perform not only a valid identification of instances but also a valid segmentation of each instance. It requires that metrics be computed both at the pixel level and at the instance level.

\begin{figure}[ht]  
\begin{subfigure}{0.48\textwidth}
\includegraphics[width=\linewidth]{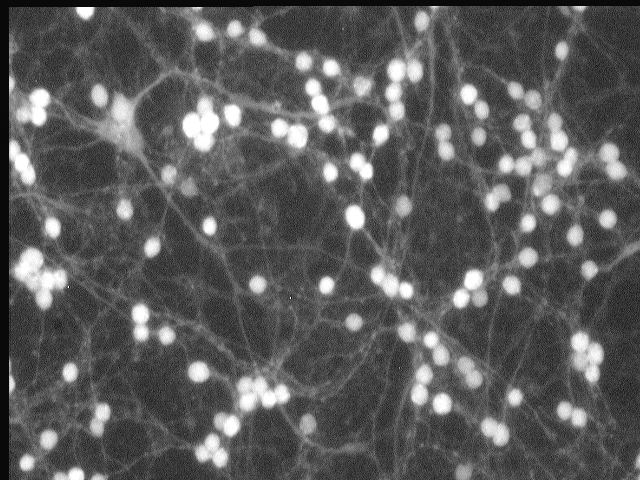}
\caption{Image 1 in fluorescence}
\end{subfigure}
\hfill 
\begin{subfigure}{0.48\textwidth}
\includegraphics[width=\linewidth]{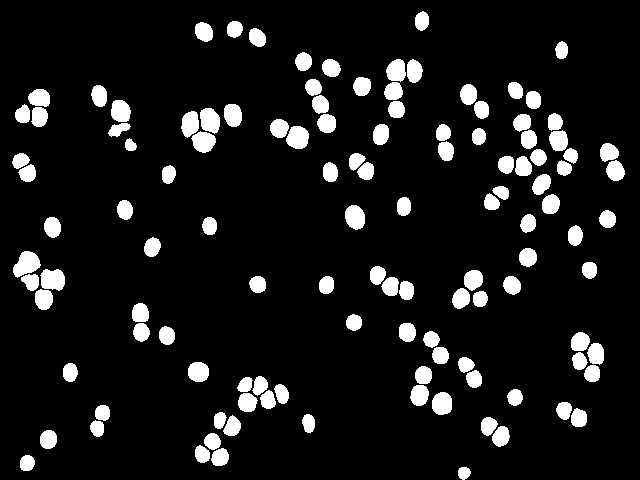}
\caption{Mask from Image 1}
\end{subfigure}

\bigskip  
\begin{subfigure}{0.48\textwidth}
\includegraphics[width=\linewidth]{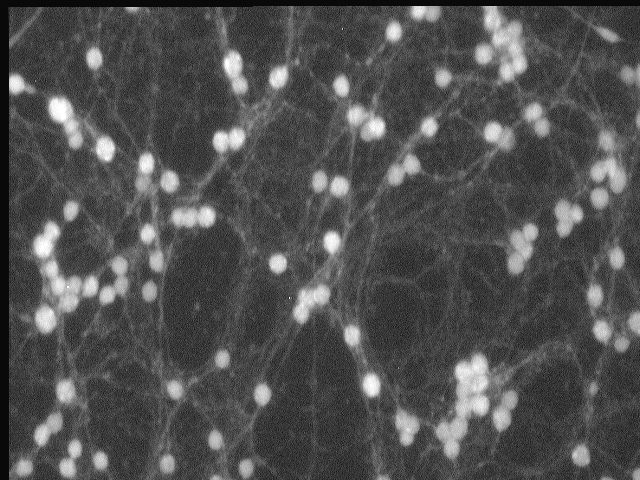}
\caption{Image 13 in fluorescence}
\end{subfigure}
\hfill 
\begin{subfigure}{0.48\textwidth}
\includegraphics[width=\linewidth]{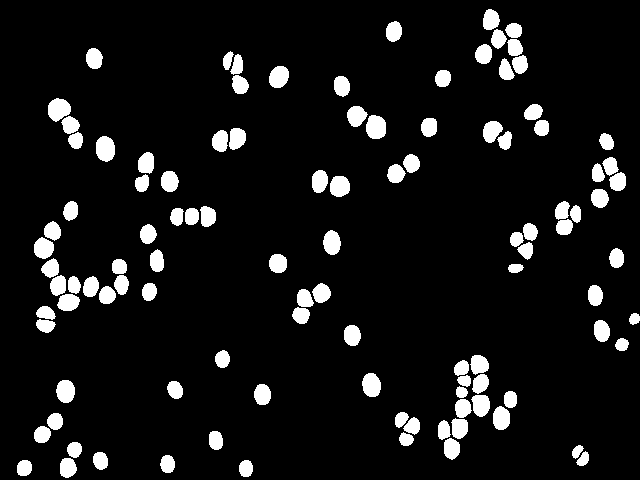}
\caption{Mask from Image 13}
\end{subfigure}

\caption{Masks generated by the algorithm from fluorescence images.}
\label{fig: F2MaskExamples}
\end{figure}

In the plots showing the results, we use the notation RX to name the Rater or Expert X. At the beginning of the analysis, the experts are ordered in no particular way, and assigned a number. The algorithm is named ``Alg" throughout the section.

\begin{figure}[ht]  
\begin{subfigure}{0.48\textwidth}
\includegraphics[width=\linewidth]{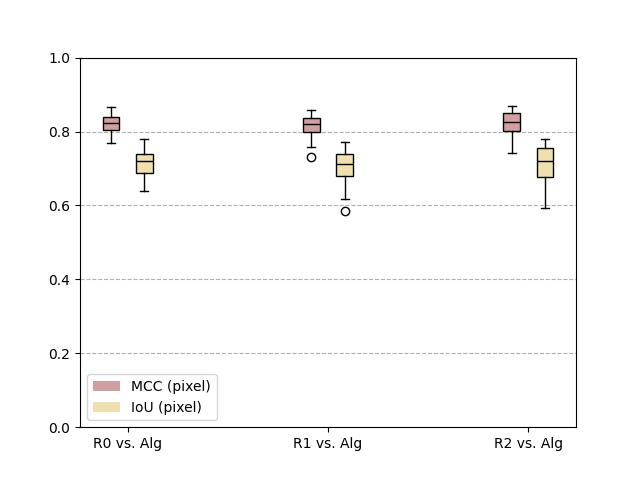}
\caption{Pixel-level metrics}
\end{subfigure}
\hfill 
\begin{subfigure}{0.48\textwidth}
\includegraphics[width=\linewidth]{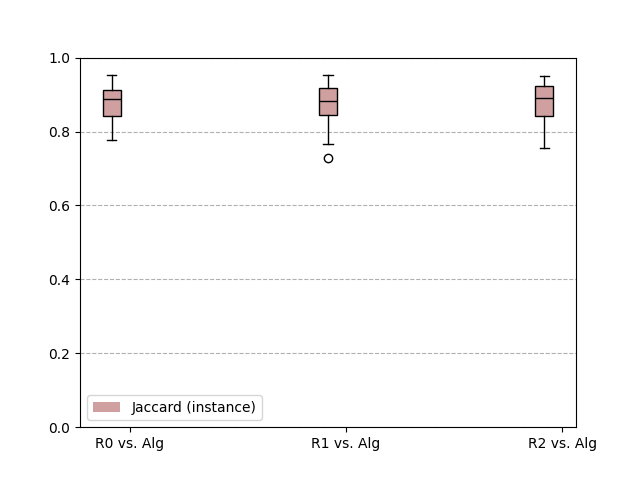}
\caption{Instance-level metrics at $IoU = 0.3$}
\end{subfigure}
\caption{Plots showing the metrics obtained from the dataset.}
\label{fig: Metric Plots}
\end{figure}
Figure \ref{fig: F2MaskExamples} shows two examples of the generated masks. The images on the left correspond to two fluorescence images as input data. The images were taken in gray scale, where bright areas correspond to the location where the flourescein has accumulated, i.e. mainly in live cells, corresponding to neurons, glial cells, and their prolongations. The masks generation task's goal is to isolate the neurons from all the rest. Not only the algorithm manages to do that successfully, but it also separates the individual instances of neurons. It shows very high accuracy even when processing clusters of neurons. There is a high visual correlation between the left and right subfigures of Figure \ref{fig: F2MaskExamples}. This correlation is quantitatively evaluated below using pixel-level and instance-level metrics. The algorithm precisely identifies the areas where neurons are located. There are cases where one neuron is segmented as two different neurons, causing a double misidentification. Conversely, some adjacent neurons are merged into a single instance.

Figure \ref{fig: Metric Plots} shows metrics at pixel and instance level. The format of the graph is a box plot, with the data grouped in quartiles. The line at the center of the plot corresponds to the median of the data and the limits of the boxes mark the medians of the upper and lower half of the data. The vertical bars indicate the maximum and minimum values in the analysis, excluding the outliers, represented by individual circles. 

Sub-figure (a) contains the MCC and the global IoU. These have been computed at a pixel level for the whole image as a whole. Although MCC is a metric that ranges from -1 to +1, the \textit{y}-axis has been limited to go from 0 to +1. Negative values aren't relevant in this study as they are not present, they indicate opposite correlation. 

The MCC in this analysis ranges from $0.73$ to $0.87$, which indicating strong agreement. The IoU metric also ranges from 0 to +1. The lower bound is reached when both masks under comparison don't share any pixels in common, and the upper bound corresponds to the situation where the masks are exactly the same. The standard value for good identification is $0.5$. Recall that the analysis at pixel level and at instance level are complementary, and thus the thresholds have to be chosen appropriately. It is clear that the algorithm performs reliably and consistently with respect to the different experts at pixel level. Such reliable metric values have been achieved for all images.

\begin{figure}[ht]  
\hfill 
\includegraphics[width=\linewidth]{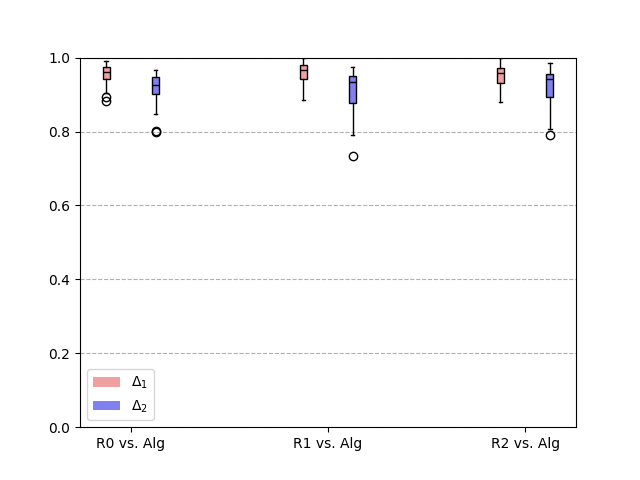}
\caption{$\Delta_1$ shows the ratio of the rater with respect to the algorithm. $\Delta_2$ is the ratio of the algorithm with respect to the rater.} 
\label{fig: Count Plots}
\end{figure}

Sub-figure (b) shows the $Jac$ coefficient at instance level. It has been computed from the confusion matrix given by the identification of neurons in both masks. As stated in the previous section, the threshold for a positive identification is fixed at $0.3$. The percentage of proper identifications with respect to all identifications into the neuron class is higher than $70\%$. The misidentification mostly come from the double counting effect mentioned in Figure \ref{fig: F2MaskExamples}, and the possible presence of artifacts in the generated mask. These artifacts can be identified in Figure \ref{fig: Count Plots}, where the accuracy of the algorithm versus the rater is lower than the rater compared against the algorithm.

The algorithm performs compatibly with an expert at pixel-level, although it consistently underperforms them slightly at instance-level. Based on the metrics and their interpretation, the segmentation the algorithm performs at pixel-level is a valid one, and the differences that arise can be assumed to be within tolerance range. 

The algorithm performs instance separation of clusters reliably. Most of neurons appear together in pairs or larger clusters. The high instance metrics, combined with the strong agreement at pixel level indicate that clusters are segmented and separated effectively. This is the main innovation of this algorithm. It performs pixel and instance classification compatible to that of the experts. Not only it processes the image based on the brightness of each pixel, but also implements shape recognition techniques successfully.

The algorithm performs instance identification whose validity is shown in Figure \ref{fig: Count Plots}. For each pair of masks we analyse, using the IoU, if each positive identified neuron in one mask has a match in the counterpart one. The process is done both ways, that is, shifting the masks. Neurons are considered correctly identified and counted towards the total number, \textit{total count}, if they have an IoU score higher than $0.3$. For each neuron that doesn't have a corresponding neuron we increase the \textit{missed neurons count}. Missed neurons also contribute to the \textit{total count}. Each mask generates a number of missed neurons: \textit{missed in mask 1} and \textit{missed in mask 2}. Both numbers are independent and they show how many of the identified neurons haven't been detected in the other mask. The sum of both counts doesn't add to the total number of missed neurons. It is greater since missed neurons appear in both counts.

To create the graph, the missed counts are divided by \textit{total count} and subtracted from 1, obtaining the ratio between mislabeled neurons and total neurons.

\begin{equation} \begin{gathered}
    \Delta_1 = 1-\frac{\textit{missed in mask 1}}{\textit{total count}}\\
    \Delta_2 = 1-\frac{\textit{missed in mask 2}}{\textit{total count}}
\end{gathered}\end{equation}

Figure \ref{fig: Count Plots} shows the area where the algorithm mostly deviates from the raters. The values that appear in the graph are due to double identification of neurons and the rare presence of artifacts caused by specific glial cells and neurites. Cluster analysis and separation of instances, although reliable, is not perfect. The median of the images are in a range close to that of the experts. The data presented here is a more detailed analysis of the sub-figure (b) in Figure \ref{fig: Metric Plots}. 

The algorithm, although consistent between raters, doesn't obtain the same metrics for all of them, which is consisten with reality as rater also show discrepancies among them. The main reason is that segmentation at pixel-level is heavily dependent on the extent of the neuron each expert includes in the mask. Different experts have an intrinsic tendency to select more or less. The identification of each instance's border is also a factor. Depending on the exact position of the drawn edge, several pixels don't line up and are counted as misidentified between both raters. All comparisons show the presence of outliers below the box. 

\section{Conclusions}\label{sec: conclusions}

This study presents a classical computer vision algorithm as an effective and fast alternative for generating neuron segmentation masks with reliable instance separation. Our results demonstrate that traditional image processing techniques can produce high-quality masks suitable for downstream applications, such as training convolutional neural networks (CNNs) for neuron detection or conducting non-invasive cell viability analyses.

While the current algorithm shows promising performance, several areas for improvement remain. The dataset used in this study is limited in size, which restricts the statistical power of the analysis. Expanding the dataset with additional expert-annotated images would enhance both robustness and generalizability. Moreover, involving a larger pool of annotators would help capture inter-expert variability, particularly in challenging cases like neuron clusters, where individual experts may differ in instance separation.

Although the algorithm can identify neuron clusters, it faces difficulties in segmenting individual neurons within dense groupings. As discussed in the results section, this does not substantially affect pixel-level metrics due to the general inclusion of clusters in the masks. However, instance-level evaluation is more sensitive to edge delineation, making precise segmentation a challenge. The algorithm performs best on high-contrast images where saturated regions are minimized, and the contrast between foreground and background supports more accurate separation.

The automatically generated masks have been used to create an expanded labeled dataset for training DL models. Our experiments show that CNNs trained on these masks perform as expected. Prior work has shown that deep neural networks are relatively robust to moderate levels of label noise  \cite{Song2020}, suggesting that the efficiency and scalability of this approach justify the trade-off in precision.

In summary, the proposed method enables the generation of expert-level neuron masks from fluorescence microscopy images, contributing to scalable, non-invasive, and DL-compatible pipelines for neurobiological research.

\section*{Acknowledgements}
This research has been funded by the Council of Gijón through the University Institute of Industrial Technology of Asturias (IUTA) grants SV-24-GIJON-1-05, SV-24-GIJON-1-18, SV-24-GIJON-1-16, SV-23-GIJON-1-09, SV-22-GIJON-1-19, and SV-21-GIJON-1-19, and by Principado de Asturias, grant SV-PA-21-AYUD/2021/50994. 

\section*{CRediT authorship contribution statement}
\textbf{Gerard Villarroya-Piqué:} Writing – review and editing, Writing – original draft, Visualization, Validation, Software, Methodology, Investigation, Formal analysis, Data curation, Conceptualization. \textbf{Víctor M. González:} Writing – review and editing, Writing – original draft, Visualization, Validation, Methodology, Investigation, Formal analysis, Data curation, Conceptualization, Project Administration, Resources, Funding acquisition. \textbf{Esther Serrano-Pertierra:} Writing – review and editing, Writing – original draft, Visualization, Validation, Methodology, Investigation, Formal analysis, Data curation, Conceptualization. \textbf{M. Teresa Fernandez-Sanchez:} Writing – review and editing, Writing – original draft, Visualization, Validation, Methodology, Investigation, Formal analysis, Data curation, Conceptualization, Resources. \textbf{Antonello Novelli:} Writing – review and editing, Writing – original draft, Visualization, Validation, Methodology, Investigation, Formal analysis, Data curation, Conceptualization, Resources. \textbf{Angel Rio-Alvarez:} Writing – review and editing, Writing – original draft, Visualization, Validation, Methodology, Investigation, Formal analysis, Data curation, Conceptualization, Project administration, Supervision.

\section*{Ethics statement}
All procedures in this research were conducted in compliance with applicable laws and institutional guidelines. The research did not require ethical approval as it was not conducted on human subjects or animals. The authors confirm that no ethical complications were identified during the course of the study.

\section*{Declaration of competing interest}
The authors declare that they have no known competing financial interests or personal relationships that could have appeared to influence the work reported in this paper.

\section*{Declaration of generative AI and AI-assisted technologies in the writing process}

During the preparation of this work the author(s) used ChatGPT and Grammarly in order to improve the readability and language of the manuscript. After using this tool/service, the authors reviewed and edited the content as needed and take full responsibility for the content of the publication.

\newpage

\bibliographystyle{unsrt}
\bibliography{bibliography}

\end{document}